\documentclass[12pt]{iopart}

\usepackage{graphicx}

\begin{document}

\title[Properties of normal and superconducting phases in the low-energy models \ldots]{Properties of normal and superconducting phases in the low-energy models of high-$T_c$ cuprates}

\author{Maxim M. Korshunov\dag \footnote[3]{To whom correspondence should be addressed (mkor@iph.krasn.ru)}, Sergei G. Ovchinnikov\dag, Alexei V. Sherman\ddag}

\address{\dag\ L.V. Kirensky Institute of Physics Siberian Branch of Russian Academy of Science, Krasnoyarsk, 660036, Russia}

\address{\ddag\ Institute of Physics, University of Tartu, Riia 142, 51014 Tartu, Estonia}

\begin{abstract}
In the framework of the effective low-energy model for High-$T_c$ cuprates with account for three-centers interaction terms and spin fluctuations the properties of normal and superconducting phases of p- and n-type cuprates are investigated. Microscopic model parameters were obtained from ARPES data on undoped compounds. Obtained evolution of the chemical potential with doping, Fermi Surface at optimal doping, and $T_c(x)$ phase diagram for n-type cuprates are in remarkably good agreement with the experiment. It is shown that the spin-exciton mechanism due to singlet-triplet hybridization takes place in p-type, although it is too small to reproduce observed qualitative difference between p- and n-type cuprates.
\end{abstract}

\pacs{74.72.Jt, 74.25.Dw, 74.25.Jb, 74.20.Mn}


\maketitle

\nosections

\textbf{1.} 
High-$T_c$ superconducting cuprates (HTSC) consist of two major classes - p-type which stands for hole-doped ($La_{2-x}Sr_xCuO_4$ - LSCO, etc.) and n-type which stands for electron-doped cuprates ($Nd_{2-x}Ce_xCuO_4$ - NCCO, $Pr_{2-x}Ce_xCuO_4$ - PCCO, etc.). Despite similar crystal structure and presence of common to all HTSC base element - $CuO_2$-plane, experimentally observed properties of these two classes are quite different. Most clear of these differences are:
i) the phase diagram \cite{Luke,Peng} is asymmetric - p-type systems characterized by the narrow range of concentrations $x$ where the antiferromagnetic (AFM) phase exist and by the wide ``dome'' of superconducting (SC) phase, while in the n-type systems the AFM phase extends until $x=0.14$ and SC phase exist only on narrow range of concentrations;
ii) distance $\Delta E_{VH}$ between the Fermi level and the Van-Hove singularity corresponding to plateau in dispersion around $(\pi,0)$ point in p-type cuprates is quite small - less then $0.03$ eV, while in NCCO $\Delta E_{VH} \approx 0.25 \div 0.35$ eV \cite{King,Armitage1};
iii) Fermi surface evolution is quite different in these types \cite{Ino1, Armitage1};
iv) until the temperature $T<100$ K the resistivity in normal state of NCCO is described by a Fermi liquid square-law dependence on temperature $T$ \cite{Hagen} in contrast to the linear dependence in p-type HTSC \cite{Takagi};
v) the insulating gap in n-type systems is indirect \cite{Armitage1} (the minimum of conduction band and maximum of a valence band are in different points of Brillouin Zone),
vi) in contrast to LSCO where pinning of the chemical potential takes place at small $x$, doping dependence of the chemical potential in NCCO is more complex \cite{Harima}.

Since the recent experimental evidence as for p-type \cite{Tsuei1} as for n-type (phase-sensitive experiments \cite{Tsuei2} and resistivity in magnetic field \cite{Panova} in NCCO, penetration depth measurements in PCCO \cite{Kokales,Prozorov}, and ARPES data \cite{Armitage2}) convincingly shows that the symmetry of the superconducting order parameter is of $d$-type (most probably, $d_{x^2-y^2}$-type) in the present work we will consider only $d_{x^2-y^2}$-pairing symmetry.

In the present work the properties of normal and superconducting phases of electron and hole doped cuprates are considered in the framework of the corresponding effective Hamiltonians for p- and n-type systems in the approximation which includes spin fluctuations beyond Hubbard-I. It is shown that this theory gives quantitative agreement with most of the experimental data on n-type compounds. In p-type systems the spin-exciton mechanism of SC pairing is investigated.

\textbf{2.}
Adequate model for High-$T_c$ cuprates is the multiband p-d model \cite{Gaididei}. Investigations of non-superconducting phase of cuprates within this model in the framework of the Generalized Tight-Binding (GTB) method with account for strong electron correlations gave quantitative agreement with ARPES data on undoped LSCO \cite{Gavrichkov1}. The chemical potential $\mu(x)$ pinning in p-type \cite{Borisov1} and absence of pinning in n-type \cite{Gavrichkov2} were described, and also the indirect insulating gap in NCCO was obtained \cite{Gavrichkov2} within GTB method. In order to investigate SC phase the low-energy effective Hamiltonian for the multiband p-d model was obtained \cite{Korshunov1} with the help of operator form of perturbation theory. Effective Hamiltonian is asymmetric for electron and hole doping - for n-type system the usual t-J model take place while for p-type systems with complicated band structure at the top of the valence band the adequate model is the effective singlet-triplet t-J model. In paper \cite{Valkov1} was shown that the influence of three-center interaction terms in the effective model could be crucial for SC phase. Finally, for n-type cuprates the effective Hamiltonian with three-center interaction terms in Hubbard X-operators representation ($X_f^{p q}=\left|p\right>\left<q\right|$) has the form:
\begin{equation} \label{model}
H_{t - J*}=H_{t - J}+H_3, \\
\end{equation}
where
\begin{eqnarray*}
\fl H_{t - J}  = \sum\limits_{f,\sigma }^{} {(\varepsilon _1-\mu) X_f^{\sigma \sigma } }  + \sum\limits_{ < f,g > ,\sigma }^{} {t_{fg}^{00} X_f^{\sigma 0} X_g^{0\sigma } }  +  
\sum\limits_{ < f,g > }^{} {J_{fg} \left( {\vec S_f \vec S_g  - \frac{1}{4}n_f n_g } \right)}, \\
\fl H_3 = \sum\limits_{ < f,g,m > ,\sigma }^{} {\frac{{t_{fm}^{0S} t_{mg}^{0S} }}{{E_{ct} }}\left( {X_f^{\sigma 0} X_m^{\bar \sigma \sigma } X_g^{0\bar \sigma }  - X_f^{\sigma 0} X_m^{\bar \sigma \bar \sigma } X_g^{0\sigma } } \right)}.
\end{eqnarray*}
Here $J_{fg} = 2\left( {t_{fg}^{0S} } \right)^2 /E_{ct} $ is the exchange integral, $E_{ct} \approx 2$ eV is the energy of charge-transfer gap that similar to $U$ in the Hubbard model, $t_{fg}^{NM}$ are the hopping parameters corresponding to annihilation of quasiparticle in state $M$ on site $f$ and creation in state $N$ on site $g$.

For p-type cuprates the effective Hamiltonian - singlet-triplet t-J model - is more complicated \cite{Korshunov1} because of beside the presence of singlet subband described by (\ref{model}) the triplet subband and the singlet-triplet hybridization are included.

All parameters of the effective Hamiltonians strictly depends on microscopic parameters of the p-d model (see paper \cite{Korshunov2,Korshunov3} where the set of microscopic and corresponding model parameters for n- and p-type cuprates are presented). Microscopic parameters for p-type (n-type) were obtained for undoped LSCO (NCCO) and in further study they are fixed and considered doping-independent. Since dependence of model parameters on distance is known from the explicit construction of Wannier states in the $CuO_2$ unit cell, the following calculations are performed with inclusion of the hoppings and exchanges up to 5-th coordination sphere.

Effective model (\ref{model}) was investigated in the framework of equation of motion method in generalized Hartree-Fock approximation \cite{Tyablikov}. In this procedure the following correlation function appears: $\left\langle {X_f^{\sigma \sigma } X_g^{\sigma '\sigma '} } \right\rangle $ and $\left\langle {X_f^{\sigma \bar \sigma } X_g^{\bar \sigma \sigma } } \right\rangle $. Hubbard I decoupling results in
\begin{eqnarray*}
\left\langle {X_f^{\sigma \sigma } X_g^{\sigma '\sigma '} } \right\rangle  \to \left\langle {X_f^{\sigma \sigma } } \right\rangle \left\langle {X_g^{\sigma '\sigma '} } \right\rangle = n_p^2, \quad
\left\langle {X_f^{\sigma \bar \sigma } X_g^{\bar \sigma \sigma } } \right\rangle  \to \left\langle {X_f^{\sigma \bar \sigma } } \right\rangle \left\langle {X_g^{\bar \sigma \sigma } } \right\rangle  = 0,
\end{eqnarray*}
where $n_p$ is the occupation factors of the one-particle state. In such decoupling the spin fluctuations are completely neglected, but the short-range magnetic order is the key element to the proper description of the properties of normal and SC phases (see e.g. \cite{Plakida,Sherman0}). Therefore, we will use decoupling which includes spin fluctuations beyond Hubbard I approximation:
\begin{eqnarray*}
\left\langle {X_f^{\sigma \sigma } X_g^{\sigma '\sigma '} } \right\rangle  \to n_p ^2 + \frac{\sigma}{\sigma'} \frac{1}{2}C_{fg}, \quad
\left\langle {X_f^{\sigma \bar \sigma } X_g^{\bar \sigma \sigma } } \right\rangle  \to C_{fg}.
\end{eqnarray*}
Here $C_{fg}  = \left\langle {X_f^{\sigma \bar \sigma } X_g^{\bar \sigma \sigma } } \right\rangle  = 2\left\langle {S_f^z S_g^z } \right\rangle $ are the spin correlation functions.

In order to calculate spin correlation functions $C_{fg}$ the two-dimensional $t-J$ model of $CuO_2$-plane was used. Self-energy equations on Green functions build with Hubbard operators were obtained with the help of Mori formalism that makes possible to present these functions as chain fraction. Elements of this fraction for electron and spin Green functions contain correlation functions for neighboring sites, while other elements of the fraction are higher-order Green functions. Latest approximated by decoupling with vertex correction \cite{Barabanov,Kondo}. Vertex correction determined from zero site magnetization restriction in considered paramagnetic case. This condition, self-energy equations on electron and spin Green functions and self-consistent conditions for correlation functions forms closed system that was iteratively solved for fixed chemical potential and temperature. Results of the calculations with small clusters \cite{Sherman1,Sherman2} are in good agreement with exact diagonalization and Quantum Monte Carlo studies. In the present work spin correlation function were calculated in described approach from spin Green function on $20 \times 20$ lattice.

In paper \cite{Sherman2} was shown that the quasiparticle decay $\Gamma _k = - {\rm Im} \Sigma _k \left( {\omega  = 0} \right)$ is large around $(0,0)$ and $(\pi,\pi)$ points. Here $\Sigma _k \left( {\omega} \right)$ is the self-energy. In further study around these points the artificial broadening of the spectral functions with $\Gamma _k$ was made. The values of $\Gamma _k$ were taken from paper \cite{Schmalian}. It is worth to mention, that our calculations shows such introduction of quasiparticle decay have very small influence on such integral characteristics as chemical potential $\mu(x)$ and SC phase transition temperature $T_c(x)$.

\begin{figure}
\begin{center}
\includegraphics[width=0.8\linewidth]{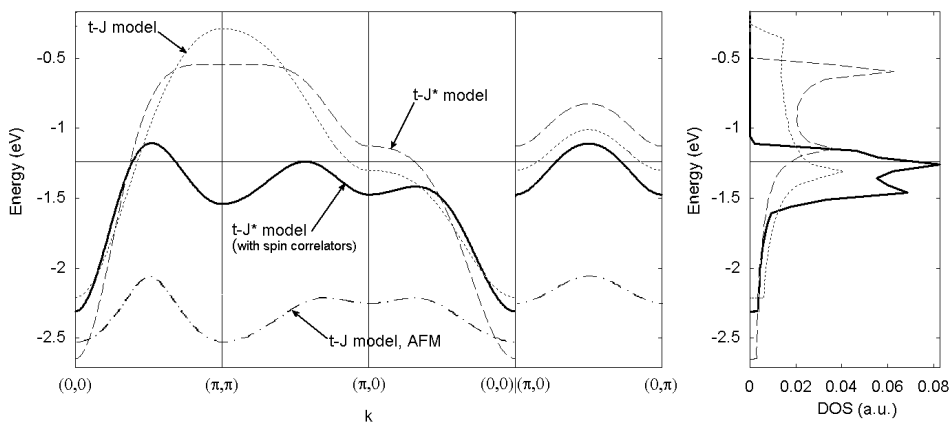}
\end{center}
\caption{\label{fig1} Dispersion curves (on the left) and density of states (on the right) in paramagnetic non-superconducting phase (optimal doping $x=0.15$) of the t-J model (dotted curve) and t-J* model (dashed curve) in Hubbard-I approximation, and of the t-J* model with account for short-range magnetic order (bold solid line). Chemical potential self-consistently calculated in the latter case denoted by solid horizontal line. Also the dispersion of the t-J model in the AFM phase is shown (dash-dotted curve).}
\end{figure}

\textbf{3.}
First, we will consider n-type cuprates and corresponding t-J* effective model (\ref{model}). Since singlet-triplet t-J* model is the generalization of the t-J* model most conclusions regarded to t-J* model will remain the same for singlet-triplet t-J* model. In Fig.~\ref{fig1} calculated dispersion curves and corresponding density of states in paramagnetic non-superconducting phase in t-J and t-J* models with and without spin fluctuations are shown. Evidently, introduction of three-center interaction terms results in significant changes on the top of the conductivity band; therefore, this will become significant at low doping. In AFM phase in t-J model there is symmetry around $(\pi/2,\pi/2)$ and $(\pi,0)$ points (see Fig.~\ref{fig1}). In paramagnetic phase this symmetry absent. But the account for spin correlation functions $C_{fg}$ gives tendency to restore symmetry around mentioned points.

\begin{figure}
\begin{center}
\includegraphics[width=\linewidth]{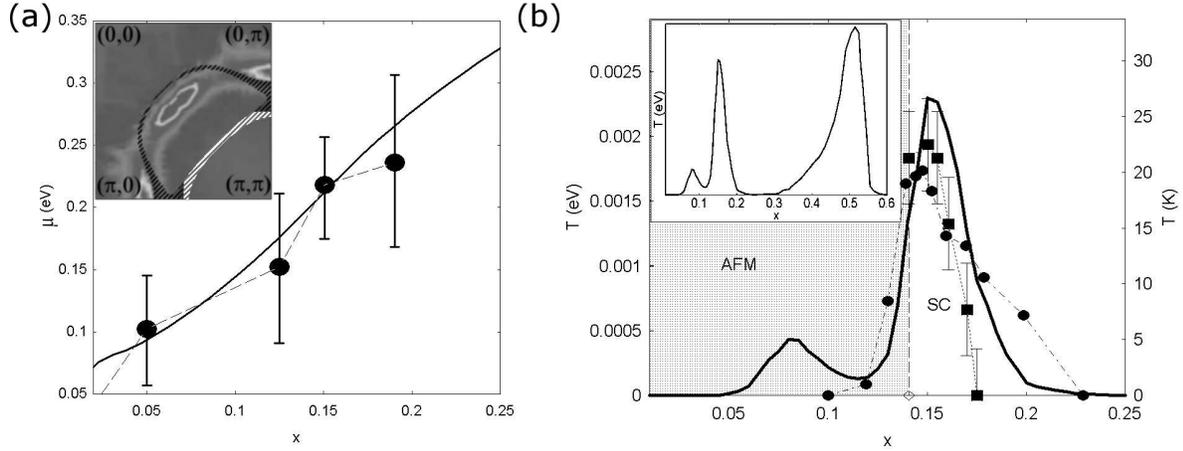}
\end{center}
\caption{\label{fig2} Results for n-type cuprates (t-J* model) with account for short-range magnetic order. (a) - calculated (solid curve) and experimental \cite{Harima} (filled circles with error bars) $\mu(x)$ dependencies. Inset: experimental \cite{Armitage3} (grayscale plot) and theoretical (black and white dashed lines, white color is used where the quasiparticle decay rate is high) Fermi surfaces of $Nd_{1.85}Ce_{0.15}CuO_4$ are presented. (b) - theoretically calculated $T_c(x)$ phase diagram (bold solid curve), and experimental results for $T_N(x)$ in NCCO (dashed line divides AFM phase on the left and paramagnetic phase on the right), $T_c(x)$ in NCCO (dotted curve with filled squares) and $T_c(x)$ in PCCO (dash-dotted curve with filled circles) are presented. Inset: calculated $T_c(x)$ dependence for wider range of concentrations.}
\end{figure}

In Fig.~\ref{fig2}a the $\mu(x)$ dependence is shown. Our theoretical calculations are in very good agreement with experimental data \cite{Harima}, presented in the same figure. In particular, the $\mu(x)$ pinning is absent. In the inset the experimental \cite{Armitage3} (grayscale plot) and calculated (black and white dashed lines) Fermi surfaces are shown for optimally doped NCCO ($x_{opt}=0.15$). ARPES experiment reveals only one Fermi surface cut and spectral function intensity is varying along this cut due to pseudogap effects. In our theory there are two Fermi surface cuts. But due to strong momentum dependence of the quasiparticle decay rate the second cut falls into area of large $\Gamma _k$ (this part shown in the inset in Fig.~\ref{fig2}a with white dashed lines). That is why this second cut must not be observed in the ARPES experiment. In the light of aforesaid, we can say that there is a good agreement of calculated and experimental Fermi surfaces.

\textbf{4.}
Since t-J* model in considered approximations give good agreements with experimental data on non-superconducting phase, now we will proceed to the SC phase investigations. Obtained in present approach equations on superconducting order parameter $\Delta_k$ are completely analogous to presented in papers \cite{Valkov1,Valkov2}, so we will not write them here. Just mention, that first, due to three-center interaction terms the SC coupling constant is strongly renormalized \cite{Valkov1}, and second, since we are taking into account hoppings and exchanges up to 5-th coordination spheres the order parameter in case of $d_{x^2-y^2}$-symmetry have the following form \cite{Valkov2}:
\begin{equation}
\Delta_k=\sum\limits_{m=1}^{2} {\Delta_m \left( \cos(m k_x a)- \cos(m k_y a) \right)}.
\end{equation}

Comparison of experimental \cite{Luke,Peng} data on NCCO and ÐÑÑÎ and our theoretical results on doping dependence of SC phase transition temperature $T_c(x)$ are presented in Fig.~\ref{fig2}b. In the inset there is the same $T_c(x)$ dependence but for wider doping range. Two maximums of $T_c$ at $x=0.15$ and $x=0.53$ are clearly seen. Also, there is very small maximum at $x=0.08$. Previously in the t-J* model with self-consistently calculated spin correlation functions the additional maximum in $T_c(x)$ for low $x$ was obtained in \cite{Valkov3}.

The picture obtained could be understood with help of simple physical argumentation. In the BCS theory there is relation between $T_c$, density of states $N\left( {\varepsilon _F } \right)$ on Fermi level $\varepsilon _F$, and effective attraction $V$: $T_c  \propto \exp \left( { - 1/N\left( {\varepsilon _F } \right)V} \right)$, ãäå $N\left( {\varepsilon _F } \right)$. Obviously, maximum in $T_c(x)$ will be gained upon coincidence of chemical potential and Van-Hove singularity. In the t-J model without spin fluctuations the only maximum in $T_c(x)$ stems from one Van-Hove singularity, corresponding to flat region in dispersion around $(\pi,0)$ point. As well known, in this case in nearest-neighbor approximation the optimal doping value is $x_{opt}=0.33$, while considering neighbors in more then three coordination spheres optimal doping become $x_{opt}=0.53$ - in the inset in Fig.~\ref{fig2}b there is one maximum $T_c(x)$ at this concentration. With inclusion of three-center interaction terms (t-J* model) but without short-range magnetic order additional singularity appears due to flat dispersion around $(\pi,\pi)$. But this point is the bottom of the conductivity band and chemical potential enters this singularity only at extremely low $x \leq 0.07$ where the mobility of carriers is low and AFM order is strong. All this leads to unfavourability of superconducting state. But in the t-J* model with proper account for spin correlation functions additional Van-Hove singularity appears at $-1.25$ eV due to saddle point around $(\pi,0.4\pi)$ (see Fig.~\ref{fig1}). This additional singularity gives maximum in $T_c(x)$ at $x \approx 0.15$. It is this singularity where the chemical potential is situated at optimal doping. It is this fact leads to the distance between position of $\mu$ and Van-Hove singularity, corresponding to plateau around $(\pi,0)$, become $\Delta E_{VH} = 0.27$ eV, that is very close to experimentally observed $\propto 0.25 \div 0.35$ eV in n-type cuprates. Note, that small maximum in $T_c(x)$ around $x=0.08$ stems from shoulder in density of states at $-1.2$ eV (see Fig.~\ref{fig1}). Furthermore, since the free energy of the AFM phase is lower then energies of normal and superconducting phases, this small maximum as well as the rest of $T_c(x)$ at concentrations lower then $x < 0.14$ and situated under experimentally obtained N\'{e}el temperature $T_N(x)$ must not be observed in experiment.

\begin{figure}
\begin{center}
\includegraphics[width=\linewidth]{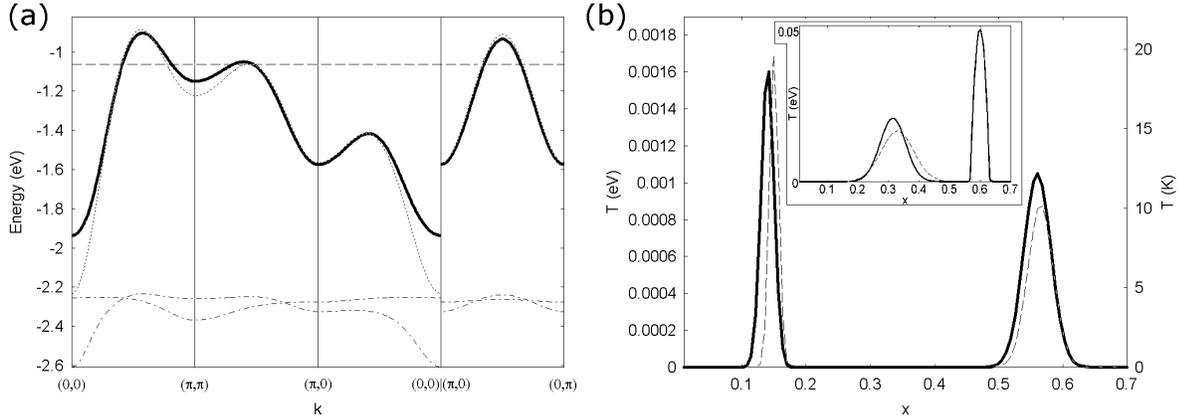}
\end{center}
\caption{\label{fig3} (a) - Comparison of the t-J* model (dotted curve) and the effective singlet-triplet t-J* model (bold solid curve - singlet subband, dash-dotted curves - triplet subbands) dispersions with spin fluctuations for optimally doped $x=0.15$ p-type cuprates. Chemical potential was self-consistently calculated in the singlet-triplet t-J* model shown by dashed line. (b) - p-type cuprates phase diagram: calculated $T_c(x)$ in the singlet-triplet t-J* model (solid bold curve) and simple t-J* model (dashed curve) are shown. In the inset the calculated $T_c(x)$ for both these models in nearest-neighbor approximation, without three-center interaction terms and without spin fluctuations is presented.}
\end{figure}

\textbf{5.}
Now we will turn to the p-type cuprates with effective singlet-triplet t-J* model. 
Comparison of dispersions in this model and t-J* model in non-superconducting phase are presented in Fig.~\ref{fig3}a.
Because in this model there are two subbands - singlet and triplet, and there is also singlet-triplet hybridization, beside SC paring in each of the subbands there could be an interband pairing. So, in p-type besides the typical for the t-J* model spin fluctuation mechanism there is additional spin-exciton mechanism of pairing \cite{Ovchinnikov1} due to singlet-triplet hybridization, and the $T_c(x)$ phase diagram for the effective singlet-triplet model is different from the phase diagram of the simple t-J model. Namely, in the nearest-neighbor approximation, neglecting three-center interaction terms and spin fluctuations,  the optimal doping value is changes from typical to t-J model value $x=0.33$ to $x=0.315$ and the maximum value of $T_c(x)$ is increased. Also, around $x=0.6$ the new ``dome'' of superconductivity appears due to triplet subband (see inset in Fig.~\ref{fig3}b). But with inclusion of three-center terms, short-range magnetic order and hoppings and exchanges up to 5-th coordination sphere, the influence of singlet-triplet induced SC pairing becomes small as seen in Fig.~\ref{fig3}b. The ``dome'' of superconductivity connected to triplet subband shifted to higher doping concentration and can't realize in experiment.

Evidently, the concurrence of spin-exciton mechanism with processes destructing SC pairs leads to constructive contribution to the $T_c(x)$ dependence. Comparison with t-J* model shows that spin-exciton mechanism leads to changes in the phase diagram - optimal doping value becomes slightly lower and the maximal $T_c(x)$ becomes slightly higher. But because the value of singlet-triple hybridization is numerically small - corresponding hopping $t^{ST} \approx 0.25 t$ (see \cite{Korshunov3}), the contribution of spin-exciton mechanism is also small. So, in present theory, the spin-exciton mechanism gives small contribution compared to dominating spin fluctuation induced SC pairing. Obtained phase diagram Fig.~\ref{fig3}b contradicts experimental one, while obtained in the same approximations phase diagram for n-type is in very good agreement with experiment Fig.~\ref{fig2}b. Also, calculated location of the Fermi level relative to Van-Hove singularity corresponding to flat region at $(\pi,0)$ is $\Delta E_{VH} = 0.52$ eV contrary to experimentally observed value $\Delta E_{VH} \le 0.03$ eV in all p-type cuprates \cite{King}. Therefore, there is another reason (beside the presence of the triplet states) for the asymmetry in p- and n-type systems to take place. Most likely, this reason connected to differences in AFM phase formation \cite{Zhang} - in n-type systems the AFM phase due to nature of doping ion could be described in diluted Heisenberg model \cite{Ovchinnikov2} while in p-type the effects of magnetic frustrations must be taken into account. Frustrations could lead to incommensurate spin fluctuations observed in inelastic neutron scattering in p-type \cite{Yamada-p} but not in n-type cuprates \cite{Yamada-n}. Also, indirect evidences of frustration's importance in p-type are spin glass state and, probably, charge separation. In our approach the non-uniform magnetic state is not taken into account, so it is justified only to n-type systems. That is why obtained in present theory phase diagram for p-type is quite different from experimentally observed.

\textbf{6.}
Summarizing, in the framework of effective model for n-type High-$T_c$ cuprates and simple physical approximations taking into account spin fluctuations beyond Hubbard I we obtained good agreement with experimental data, such as the evolution of chemical potential with doping, Fermi surface at optimally doped NCCO and $T_c(x)$ dependence. Although analogous results of $T_c(x)$ dependence were previously obtained in FLEX approximation \cite{Manske}, the presented approach is superior due to its physical transparency and the explicit account for strong electron correlations, which plays very significant role in HTSC. We have shown, that the underlying physics of $T_c(x)$ concentration dependence in n-type cuprates is quite different from p-type counterparts. Namely, as the consequence of sort-range magnetic order (and corresponding spin fluctuations) the system have tendency to restore AFM symmetry. This leads to changes in dispersion so the flat areas appears around $(\pi,0.4\pi)$ (and symmetric to it) point. Flat region produces additional Van-Hove singularity. Due to this transformation of density of states the new ``dome'' of superconductivity appears at $x$ around $0.15$. This ``dome'' is in remarkably good agreement with experimental results - not only the value of the optimal doping $x_{opt}$ and corresponding $T_c(x)$ but also the range of dopant concentrations where the superconductivity exist are reproduced. We will mention again, that in present investigation there were no fitting parameters - all parameters of effective model in strict relationship with microscopic parameters of the multiband p-d model. And these microscopic parameters were determined in our previous papers from ARPES data on undoped AFM cuprates.
So, we have formulated quantitative microscopic theory of High-$T_c$ superconductivity in electron doped cuprates. This theory properly (i.e. in agreement with the experiment) describes properties not only in superconducting but also in normal phases of HTSC.

For p-type cuprates the influence of singlet-triplet hybridization is analyzed. Due to this hybridization the spin-exciton mechanism of SC pairing appears to take place, but because of small numerical value of singlet-triplet hybridization the spin-exciton mechanism gives only very small contribution to the $T_c(x)$ dependence. We believe that the main mechanism of SC pairing in p-type systems is the spin fluctuation induced mechanism, but in order to explain complicated phase diagram of these compounds one must take into account magnetic frustrations, that takes place in underdoped region and leads to non-uniform magnetic state, incommensurability of spin fluctuations and phase separation observed in hole doped cuprates.

Concerning the electron-phonon interaction which was not taken into account in the present work, there are strong indications of weakness of this interaction in n-type cuprates. First, the isotope effect is almost absent in these compounds \cite{Kulic}. Second, there is no kink feature along $(0,0)-(\pi,\pi)$ direction \cite{Armitage4} that is significantly different from p-type where appearance of kink considered as evidence for strong electron-phonon interaction. But in hole doped cuprates there could be additional contribution due to this interaction \cite{Nunner}.

\begin{ack}
The authors are thankful to V.V. Val'cov and D.M. Dzebisashvili for the stimulating discussion. This work was supported by RFBR grant 03-02-1624, Russian Academy of Science Program ``Quantum Macrophysics'', INTAS grant 01-0654, ETF grant 5548, Siberian Branch of RAS (Lavrent'yev Contest for Young Scientists), The Dynasty Foundation and ICFPM.
\end{ack}

\end{document}